\documentclass[twocolumn]{aa501}
\usepackage{psfig} 

\def\kms{$\mathrm{km\;s}^{-1}$} 
\def\dg{^\circ} 
\def\ha{H$\alpha$}
\def\h2{H$_{2}$}

\def\nii{[N~{\small II}]}

\def\niipg{[N~{\small II}]$\,\lambda\lambda6548,6583$}

\def\siipg{[S~{\small II}]$\,\lambda\lambda6716,6731$}

\begin{document} 
 
\title{The Orthogonal Gaseous Kinematical Decoupling in the Sa Spiral
  NGC 2855
\thanks{Based on observations carried out at the Observatorio 
del Roque de los Muchachos, La Palma (Spain) with the Italian
Telescopio Nazionale Galileo and at the European Southern Observatory,
La Silla (Chile) (ESO 62.N-0463 and 67.B-0230).}$^{\bf,}$
\thanks{Tables 3 to 4 are only available in electronic form at the 
CDS via anonymous ftp to cdsarc.u-strasbg.fr (130.79.128.5) or via
http://cdsweb.u-strasbg.fr/Abstract.html.}}

 \author{ 
     E.M.~Corsini           \inst{1},  
     A.~Pizzella            \inst{2}, 
     and F.~Bertola         \inst{2}}

\offprints{Enrico Maria Corsini,\\ email:{\tt corsini@pd.astro.it}}

\institute{
Osservatorio Astrofisico di Asiago, Dipartimento di Astronomia, 
  Universit\`a di Padova, via dell'Osservatorio~8, I-36012 Asiago, Italy \and  
Dipartimento di Astronomia, Universit\`a di Padova, 
  vicolo dell'Osservatorio~2, I-35122 Padova, Italy} 

\date{Received September 2, 2001/ Accepted October 26, 2001}  
  
\titlerunning{NGC 2855}  
\authorrunning{Corsini, Pizzella, and Bertola}

\abstract{We present major and minor-axis kinematics of stars and
  ionized gas as well as narrow and broad-band surface photometry of
  the Sa spiral NGC 2855.  In the nuclear regions of this unbarred and
  apparently undisturbed spiral galaxy the gas is rotating
  perpendicularly to the galaxy disk. We suggest that this
  kinematically-decoupled component is the signature of an acquisition
  process in the history of this galaxy.
  }

\maketitle

%%%%%%%%%%%%%%% 
% INTRODUCTION 
%%%%%%%%%%%%%%% 
\section{Introduction} 
\label{sec:n2855_introduction} 

Like fossils, multiple stellar and gaseous components with a
misaligned or even opposite angular momentum with respect to the host
galaxy conserve memory of processes driving galaxy formation and
evolution. In disk galaxies apart from polar rings they usually lurk
to morphological inspection and are serendipitously discovered by
measuring detailed kinematics (see Bertola \& Corsini 1998 for a
review).
This the case of a variety of kinematically-decoupled components
ranging from the counterrotating disks of ionized gas observed in a
large fraction of S0's (Bertola, Buson \& Zeilinger 1992; Kuijken,
Fisher \& Merrifield 1996) as well as in one Sa, \object{NGC 3626}
(Ciri, Bettoni \& Galletta 1995; Haynes et al. 2000), to the two
counterrotating stellar disks found in the S0 \object{NGC 4550} (Rubin
et al. 1992; Rix et al. 1992), in the early-type spirals \object{NGC
  3593} (Bertola et al. 1996), \object{NGC 4138} (Jore, Broeils \&
Haynes 1996; Haynes et al. 2000), and \object{NGC 7217} (Merrifield \&
Kuijken 1994), to the orthogonally-rotating stellar cores discovered
in two Sa spirals, namely \object{NGC 4698} (Bertola et al. 1999) and
\object{NGC 4672} (Sarzi et al. 2000), and to the two counterrotating
gaseous components as in the S0 \object{NGC 7332} (Fisher, Illingworth
\& Franx 1994; Plana \& Boulesteix 1996) and in the Sab \object{NGC
  4826} (Braun et al. 1992, 1994; Rubin 1994).

The recent findings by Bettoni et al. (2001) on the higher gas content
of counterrotators and polar ring galaxies with respect to normal
galaxies gives further support to the idea that
kinematically-decoupled components are the end result of one or more
second events occurred in the history of the host. Yet, even with this
insight, the undisturbed appearance with no evident signatures of
interaction characterizing most of galaxies with kinematic
peculiarities (Bettoni, Galletta \& Prada 2001) poses questions on the
effective rate of second events and on their role in determining
Hubble types.
Accretion of external gas and minor mergers seem to be the favoured
mechanisms to form decoupled components preserving the pre-existent
disk (e.g. Thakar \& Ryder 1996; Thakar et al. 1997), nevertheless
major mergers can not be ignored since for a narrow range of initial
conditions they still represent a viable alternative in building
massive counterrotating disks (Pfenniger 1999) and polar ring galaxies
(Bekki 1998).
Bulge growing, disk heating and spiral-pattern smoothing are among the
observable consequences of second events in disk galaxies, as
predicted by numerical experiments (e.g. Aguerri, Balcells \& Peletier
2001; Quinn, Hernquist \& Fullagar 1993; Thakar \& Ryder 1998). They
result in a change of the host morphology toward earlier spiral
types.
 
Recently it has been reported the presence of a velocity gradient
along the minor axis of the disk in the innermost region of two Sa
spirals, namely NGC 4698 (Bertola et al. 1999) and NGC 4672 (Sarzi et
al. 2000), characterized by an uncommon and remarkable orthogonal
geometrical decoupling between bulge and disk.  In NGC 4698 the
minor-axis velocity gradient is observed in both stars and ionized
gas, while in NGC 4672 only in the stellar component.  Such a stellar
rotation along the disk minor axis indicates the presence of a
kinematically isolated core, which is rotating perpendicularly with
respect to the disk component. The analysis of the HST images of the
nucleus of NGC 4698 shows that in this case isolated core is a nuclear
stellar disk with a scalelength of few tens of pc (Pizzella et al.
2001).
According to Bertola \& Corsini (2000) the presence of these
orthogonally-rotating isolated cores indicates that the entire disk of
the galaxy could be the end result of the acquisition of external
material in polar orbits around a pre-existing oblate spheroid, which
became the bulge of the present galaxy. In this scenario the isolated
core has been formed by the gaseous material settled down in the
symmetry plane of the oblate spheroid during the acquisition process.

In this paper we present a new case of kinematical orthogonal
decoupling in the bulge of early-type spiral galaxies which
characterizes the gaseous component only. In fact the ionized gas
observed in the innermost regions ($|r|\la0.4$ kpc) of the Sa
\object{NGC 2855} is in orthogonal rotation with respect to the rest
of the galaxy.

The paper is organized as follows. We present our photometric and
spectroscopic observations of NGC 2855 in Sect.
\ref{sec:n2855_observations} and we analyze its morphological and
kinematical properties in Sects. \ref{sec:n2855_morphology} and
\ref{sec:n2855_kinematics}, respectively. Our conclusions are
discussed in Sect.  \ref{sec:n2855_conclusions}.

%%%%%%%%%%%%%%%%%%%%%%%%%%%%%%%%%% 
% OBSERVATIONS AND DATA REDUCTION
%%%%%%%%%%%%%%%%%%%%%%%%%%%%%%%%%% 
\section{Observations and data reduction} 
\label{sec:n2855_observations} 

\subsection{Narrow and broad-band imaging}

The narrow and broad-band imaging of NGC 2855 were carried out at the
Roque de los Muchachos Observatory in La Palma with the Telescopio
Nazionale Galileo (TNG) on April 18, 1999 and at the European Southern
Observatory (ESO) in La Silla with the 3.6-m ESO telescope on March
27, 2001. The instrumental setup and the observing log are collected
in Table~\ref{tab:photometry}.

\begin{table}[!ht]
\caption{Setup and log of photometric observations.}
\begin{tabular}{lccc}
\hline
\noalign{\smallskip}
Parameter  & April 18, 1999 & \multicolumn{2}{c}{March 27, 2001} \\
\noalign{\smallskip}
\hline 
\noalign{\smallskip}  
Instrument    & TNG$+$OIG           & 
 \multicolumn{2}{c}{ESO 3.6-m$+$EFOSC2} \\ 
CCD           & 2$\times$EEV 42-80  & 
 \multicolumn{2}{c}{\#40 Loral/Lesser} \\
Pixel size    & $13.5\times13.5$ $\mathrm{\mu\;m}^2$ &
  \multicolumn{2}{c}{$15\times15$ $\mathrm{\mu\;m}^2$} \\
Scale         & $\mathrm{0\farcs14\;pixel}^{-1}$ &
  \multicolumn{2}{c}{$\mathrm{0\farcs31\;pixel}^{-1}$} \\
Field of view & $4\farcm9\times4\farcm9$ & 
  \multicolumn{2}{c}{$5\farcm3\times5\farcm3$}\\
Filter        & $R$                 & \#438 \ha$^{\rm a}$ & \#642 $R$ \\
Exposure time &  $2\times300$ s     & $2\times600$ s     & $2\times150$ s \\
Seeing FWHM$^{\rm b}$ &  $2\farcs1$ & $0\farcs9$          & $0\farcs9$ \\ 
\noalign{\smallskip}  
\hline 
\noalign{\smallskip}  
%\end{footnotesize}
\end{tabular}\\  
\begin{minipage}{8.8cm} 
$^{\rm a}$ $\lambda_c = 6630$ \AA, $\Delta\lambda_{\rm FWHM} = 68$ \AA.\\
$^{\rm b}$ Measured on averaged frames.
\end{minipage}
\label{tab:photometry}
\end{table}

\begin{figure}[ht] 
\centerline{\hbox{ 
     \psfig{figure=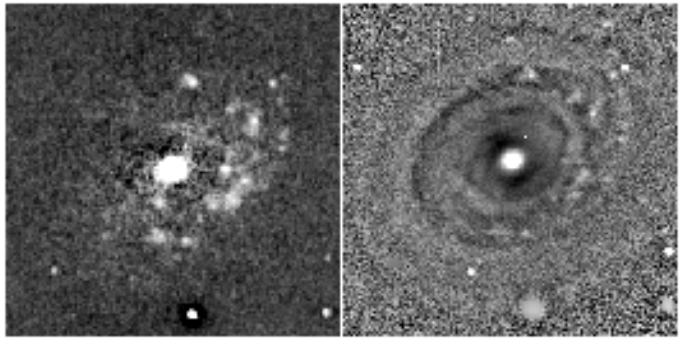,width=9cm}}}
\caption[]{Continuum-subtracted \ha$+$\nii\ 
  emission image ({\it left panel\/}) and unsharp masking version of
  the narrow-band image ({\it right panel\/}) of NGC 2855 unveiling
  the spiral structure of the galaxy.  The size of images is
  $1'\times1'$. North is up and east is left.}
\label{fig:halpha_unsharp} 
\end{figure}

Using standard MIDAS\footnote{MIDAS is developed and maintained by the
  European Southern Observatory} routines the images were bias
subtracted, corrected for flatfield using sky flats taken at the
beginning and end of each observing night, and cleaned for cosmic
rays. The sky background level was removed by fitting a second order
polynomial to the regions free of sources in the images.  The images
obtained with the same filter at the same telescope were shifted and
aligned using the common field stars and then they were averaged
(after checking their PSF's were comparable). The map of the
\ha$+$\nii\ emission was obtained by subtracting the ESO $R-$band
image, suitably scaled, from the narrow-band image (Fig.
\ref{fig:halpha_unsharp}).
To confirm the spiral classification of NGC 2855 we built an
unsharp-masked image of the galaxy, by dividing its narrow-band frame
by itself after being convolved with a circular two-dimensional
Gaussian of $\sigma\,=\,1\farcs2$ (Fig. \ref{fig:halpha_unsharp}).
This procedure enhanced the spiral pattern of the galaxy as well as
any surface-brightness fluctuation and non-circular structure
extending over a spatial region comparable with the size of the
smoothing Gaussian.
Absolute calibration was performed only for the $R-$band images.

\begin{figure}[ht] 
\centerline{\hbox{ 
     \psfig{figure=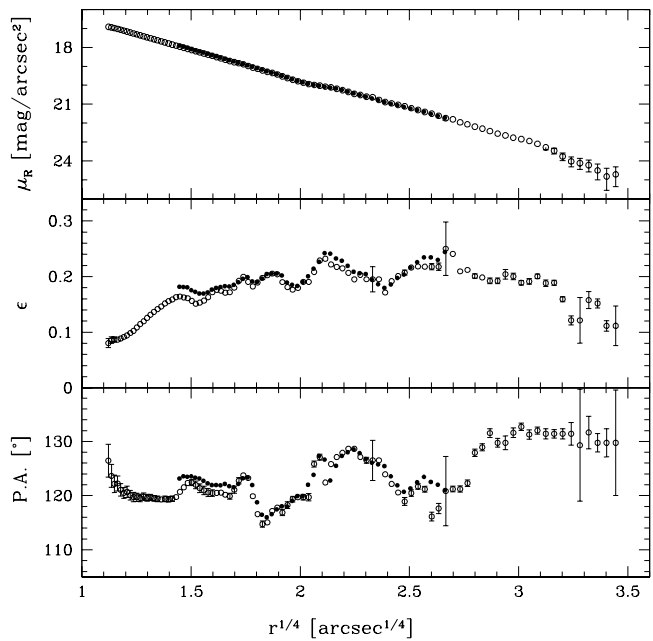,width=9cm}}}
\caption{$R-$band surface-brightness, ellipticity and position angle
  radial profiles of NGC 2855 measured on the TNG ({\it filled
  circles\/}) and 3.6-m ({\it open circles\/}) images.  Errorbars
  smaller than symbols are not plotted.}
\label{fig:photometry}
\end{figure}

We analyzed the isophotal profiles of the galaxy by masking the field
stars and fitting ellipses to the isophotes using the
IRAF\footnote{IRAF~is distributed by NOAO, which is operated by AURA
  Inc., under contract with the National Science Foundation} task
ELLIPSE. We first allowed the centers of the ellipses to vary, to test
whether the optical disk is disturbed.  Within the errors of the fits,
we found no evidence of a varying center, therefore the ellipse fits
were therefore repeated with the ellipse centers fixed.  The resulting
surface brightness, ellipticity and position angle radial profiles
obtained from both the ESO and TNG $R-$band images are given in Fig.
\ref{fig:photometry}.

\subsection{Long-slit spectroscopy}

The spectroscopic observations of NGC 2855 were carried out at the ESO
in La Silla with the 1.52-m ESO telescope on March 17--22, 1999 and
New Technology Telescope (NTT) on May 22, 2001. The details of the
instrumental setup and the log of these observing runs are given in
Table~\ref{tab:spectroscopy}.

\begin{table*}[!ht]
\caption{Setup and log of spectroscopic observations.}
\begin{small} 
\begin{tabular}{lccc} 
\hline
\noalign{\smallskip}
Parameter  & \multicolumn{2}{c}{March 17--22, 1999} & 
 May 22, 2001 \\
\noalign{\smallskip}
\hline 
\noalign{\smallskip}  
Instrument   & \multicolumn{2}{c}{ESO 1.52-m$+$B\&C}           & 
 NTT$+$EMMI \\ 
Grating/Grism& \multicolumn{2}{c}{\#33 1200 $\mathrm{gr\;mm}^{-1}$} & 
 \#7 600 $\mathrm{gr\;mm}^{-1}$ \\
CCD          & \multicolumn{2}{c}{\#39 Loral/Lesser}           & 
 \#36 TEK 2048 \\
Pixel size   & \multicolumn{2}{c}{$15\times15$ $\mathrm{\mu\;m}^2$} &
 $24\times24$ $\mathrm \mu\;m^2$ \\
Scale        & \multicolumn{2}{c}{$\mathrm{0\farcs82\;pixel}^{-1}$} & 
 $\mathrm{0\farcs27\;pixel}^{-1}$ \\ 
Dispersion   &  \multicolumn{2}{c}{$\mathrm{0.98\;\AA\;pixel}^{-1}$} & 
 $\mathrm{0.65\;\AA\;pixel}^{-1}$ \\ 
Slit width   & \multicolumn{2}{c}{$2\farcs2$}                  & 
 $1\farcs0$ \\ 
Wavelength range       & \multicolumn{2}{c}{4812 -- 6824 \AA}  & 
 5630 -- 6968 \AA \\ 
Instrumental $\sigma^{\mathrm a}$  &  \multicolumn{2}{c}{$52$ \kms}& 
 $28$ \kms \\ 
Position angle         &  $30\dg$         & $120\dg$           & 
 $30\dg$ \\
Exposure time          &  $3\times60$ min & $2\times60$ min    & 
 $30$ min \\
Seeing FWHM  &  \multicolumn{2}{c}{$1\farcs5$ -- $2\farcs5$}   & 
 $0\farcs8$ \\ 
\noalign{\smallskip}  
\hline 
\noalign{\smallskip}  
\end{tabular}\\  
\begin{minipage}{8.8cm} 
$^{\mathrm a}$ Measured at \ha .
\end{minipage} 
\end{small}
\label{tab:spectroscopy}
\end{table*}

At the beginning of each exposure the galaxy was centered on the slit
using the guiding camera.  During the March 1999 observing run, we
took spectra of several giant stars whose spectral type ranged from
G5III to K7III to be used as templates in measuring the stellar
kinematics. Comparison spectra were taken before every object
exposure.
The reduction was carried on with MIDAS. All the spectra were bias
subtracted, flatfield corrected, and wavelength calibrated. The
contribution of the sky was determined from the outermost $10''$ at
the two edges of the resulting spectra, where the galaxy light was
negligible, and then subtracted. The spectra taken along the same axis
in the same run were co-added using the center of the stellar
continuum as reference, and cosmic rays have been cleaned up in the
process.

\begin{figure*}[ht!] 
% no psfig used to avoid problems with astro-ph submission
\vspace*{16cm} 
\includegraphics{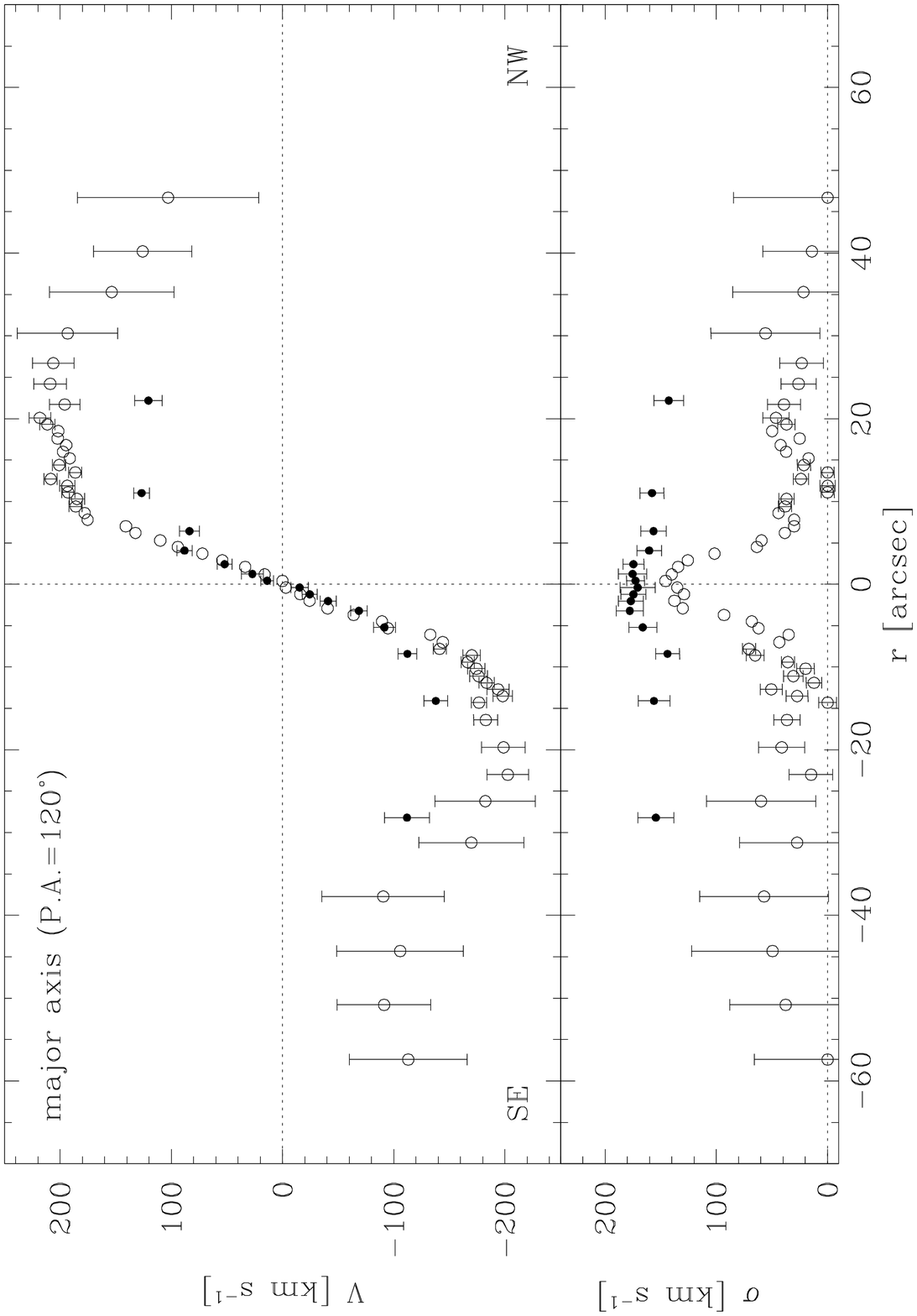}
\includegraphics{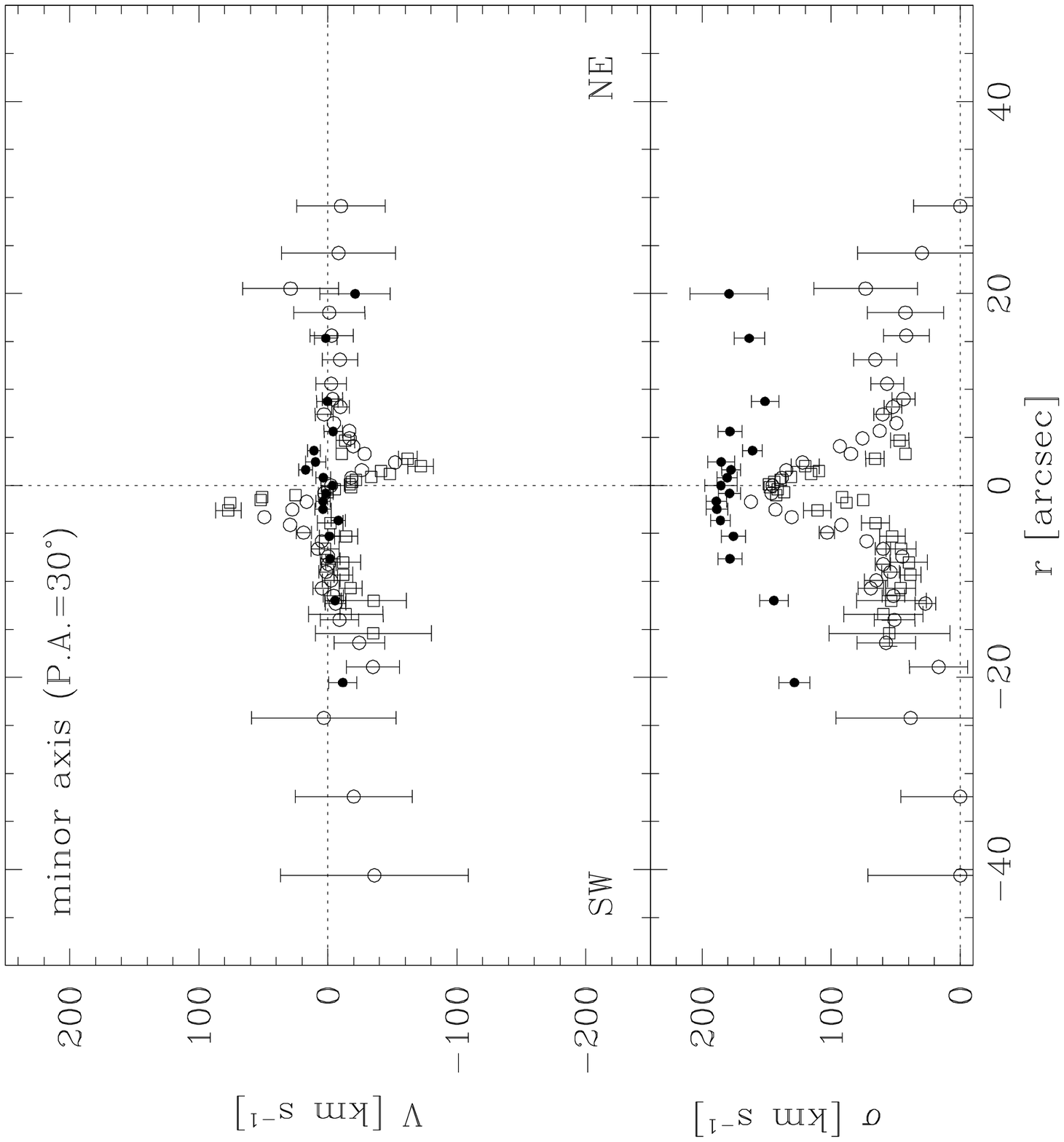}
\caption[]{The stellar ({\it filled circles\/}) and ionized-gas
  kinematics ({\it open symbols\/}) measured along the major ({\it
    upper panel\/}) and minor axis ({\it lower panel\/}) of NGC 2855.
  All velocities are plotted as observed after subtracting the value
  of the systemic heliocentric velocity and without applying any
  inclination correction.  Errorbars smaller than symbols are not
  plotted. In the lower panel {\it open circles\/} and {\it open
    squares\/} represent data obtained with the ESO 1.52-m telescope
  and NTT, respectively.}
\label{fig:kinematics}
\end{figure*}

We measured the stellar kinematics from the galaxy absorption features
present in the wavelength range running from 5064 \AA\ to 5540 \AA\ 
and centered on the Mg line triplet ($\lambda\lambda\,5164,5173,5184$
\AA).
We used the Fourier Correlation Quotient method (FCQ, Bender 1990)
following the prescriptions of Bender, Saglia \& Gerhard (1994).  The
spectra were rebinned along the spatial direction to obtain a nearly
constant signal-to-noise ratio larger than 20 per resolution element
(with a peak of 80 in the innermost regions of the spectra).  The
galaxy continuum was removed row-by-row by fitting a fourth to sixth
order polynomial as in Bender et al.  (1994). \object{HR 2035} was
adopted as kinematical template. This allowed us to derive, for each
spectrum, the line-of-sight velocity distribution (LOSVD) along the
slit and to measure its moments, namely the radial velocity $v$, the
velocity dispersion $\sigma$ and the values of the coefficients $h_3$
and $h_4$.  At each radius, they have been derived by fitting the
LOSVD with a Gaussian plus third- and fourth-order Gauss-Hermite
polynomials $H_3$ and $H_4$, which describe the asymmetric and
symmetric deviations of the LOSVD from a pure Gaussian profile (van
der Marel \& Franx 1993; Gerhard 1993).
We derived errors on the LOSVD moments from photon statistics and CCD
read-out noise, calibrating them by Monte Carlo simulations as done by
Bender et al. (1994). These errors do not take into account possible
systematic effects due to any template mismatch.  The resulting
stellar kinematics is reported in Table 3 and plotted in Fig.
\ref{fig:kinematics}.

The ionized-gas kinematics was measured by the simultaneous Gaussian
fit of the emission lines present in the spectra (namely \niipg, \ha,
and \siipg) with an `ad hoc' procedure written within the IDL
environment.  The galaxy continuum was removed from the spectra as
done for measuring the stellar kinematics. We fitted in each row of
the continuum-subtracted spectrum a Gaussian to each emission line,
assuming them to have the same line-of-sight velocity and velocity
dispersion (corrected for heliocentric velocity and instrumental FWHM,
respectively). An additional absorption Gaussian has been added in the
fit to take into account for the presence of the \ha\ absorption line
and the flux ratio of the \niipg\ lines have been fixed to 1:3. Far
from the galaxy center (for $|r|\ga10''$) we averaged adjacent
spectral rows in order to increase the signal-to-noise ratio of the
relevant emission lines.
We calibrated the formal velocity errors obtained from the
least-squares fit to match those derived by measuring the rotation
curve of night-sky emission lines (cf. Corsini et al. 1999).
The resulting ionized-gas kinematics is given in Table 4 and show in
Fig. \ref{fig:kinematics}.

%%%%%%%%%%%%%
% MORPHOLOGY
%%%%%%%%%%%%%
\section{Morphology of NGC 2855} 
\label{sec:n2855_morphology} 
 
NGC~2855 is an early-type spiral galaxy classified as
Sa(r) by Sandage \& Tammann (1981, RSA hereafter) and RS0/a(rs) by de
Vaucouleurs et al. (1991, RC3 hereafter).  
Its total $B-$band magnitude is $B_T=12.63$ mag (RC3) which
corresponds to $M_B=-19.11$ mag for a distance $D=22.3$ Mpc.  We
derive the distance from the heliocentric systemic velocity we
measured ($V_\odot = 1897\pm17$ \kms) assuming $H_0 = 75$ \kms\ 
Mpc$^{-1}$ and following the precepts of RSA.  The measured systemic
velocity is in agreement within the errors with the values listed by
RSA, Tully (1988) and RC3. The inclination $i=39\dg$ was obtained from
the observed axial ratio (whose mean value between $15''$ and $30''$
is $q=0.79$, Fig. \ref{fig:photometry}) after correcting for the
typical intrinsic flattening of S0/a galaxies ($q_0 = 0.18$, Guthrie
1992).

Sandage (1961) in The Hubble Atlas of Galaxies considered the galaxy
as transition case from S0$_3$ to Sa(r) due to its circular central
absorption ring. Sandage \& Bedke (1994) included NGC 2855 in the
section of Carnegie Atlas of Galaxies devoted to Sa galaxies with a
spiral pattern entirely defined by dust lanes (Panels 73 and 74).  The
presence of tightly wounded spiral arms in the disk is confirmed by
our continuum-subtracted \ha$+$\nii\ emission image of NGC 2855 and by
the unsharp-masked version of its narrow-band image (Fig.
\ref{fig:halpha_unsharp}).

\begin{figure*}[ht] 
\centerline{\hbox{ 
     \psfig{figure=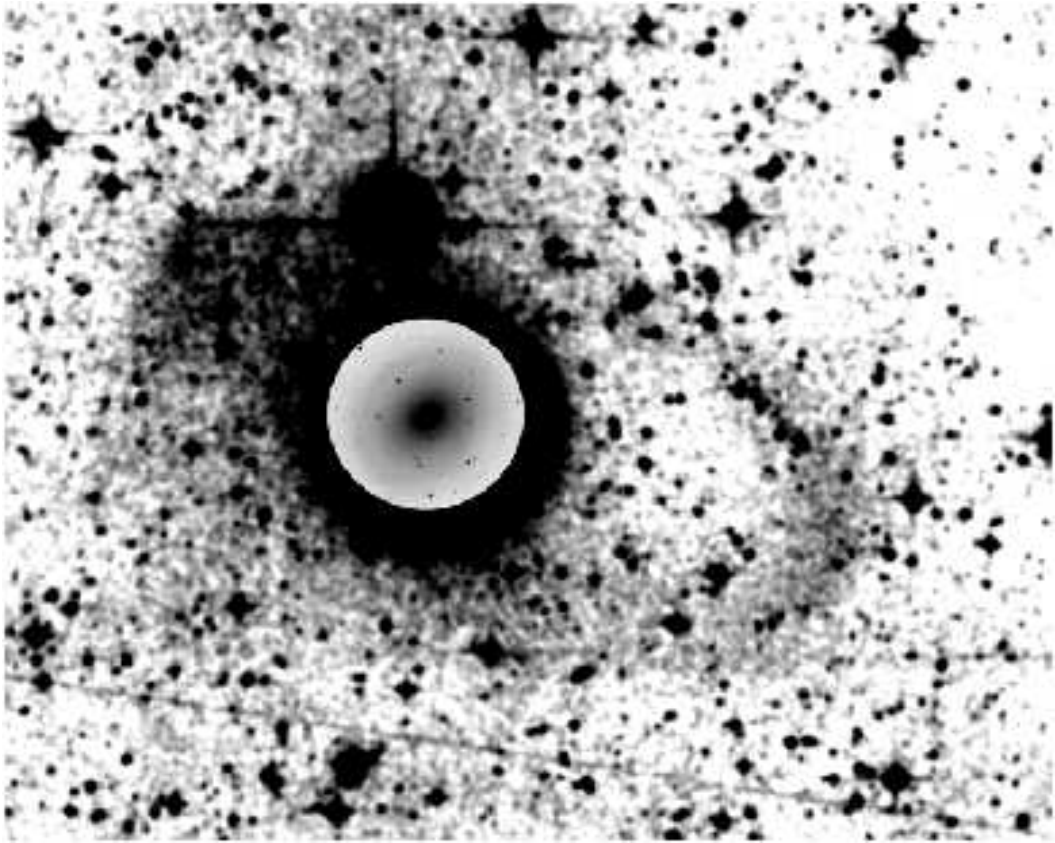,width=14cm}}}
\caption{Deep image of the NGC 2855 field adapted from Malin \& Hadley (1997)
  and processed to enhance the ring structure surrounding the galaxy.
  Our ESO $R-$band image is shown in the inset. The frame is $15'$
  height. North is up and east is left.}
\label{fig:malin}
\end{figure*}

The morphology NGC 2855 appears relatively undisturbed (Rudnick \& Rix
1998) and the galaxy has no close companions (RC3) suggestive of
evident interaction phenomena.  However enhanced IIIa-J plates with
deep exposures of the NGC 2855 field (Malin \& Hadley 1997) reveal the
presence of a structure of low surface brightness ($\ga28$ mag
arcsec$^{-2}$), which surrounds the galaxy. We processed this image,
which was kindly put at our disposal by D. Malin, to enhance the
contrast of the faint structure (Fig. \ref{fig:malin}). We found that
this almost complete loop is elongated in a direction close to the
galaxy minor axis ($\mathrm{P.A.}\,\simeq\,60\dg$) and it extends out
to about $6'$ and $8'$ from the galaxy center on its NE and SW side,
respectively. It resembles an offset ring with an inclination of about
$70\dg$.  At fainter levels the stellar disk is nearly face on,
confirming a trend already present in the ellipticity profile shown in
Fig. \ref{fig:photometry}. Indeed we derive an inclination $i=27\dg$
at $r=141''$, corresponding to the outermost observed radius in our
ESO $R-$band image. This change of inclination is associated to a
variation of the isophotal position angle
($\mathrm{P.A.}\,\simeq\,130\dg$ between $53''$ and $141''$, and
$\mathrm{P.A.}\,\simeq\,40\dg$ at $2\farcm6$ from Fig.
\ref{fig:malin}), and is indicative of a warped stellar disk.
The large extension and possible warp of the disk component could
explain the discrepancy between the available bulge-disk
decompositions of the surface brightness of NGC 2855. Indeed the disk
results embedded in the bulge at all the observed radii according to
the $V$ (Baggett, Baggett \& Anderson 1998) and $K-$band photometric
decompositions (M\"ollenhoff \& Heidt 2001). On the contrary the disk
dominates the galaxy light for $r\ga35''$ in the $B$ and $J-$band
photometric decomposition performed by Boroson (1981) and M\"ollenhoff
\& Heidt (2001), respectively.

%%%%%%%%%%%%%
% KINEMATICS
%%%%%%%%%%%%%
\section{Kinematics of NGC 2855} 
\label{sec:n2855_kinematics} 
 
The velocity curves and velocity dispersion profiles we measured for
the stellar and gaseous components along the major and minor axis of
NGC 2855 are presented in Fig. \ref{fig:kinematics}.

The stellar kinematics does not show any peculiarity on both the
observed axes. It extends out to $28''$ (corresponding to $3.0$ kpc)
and $20''$ from the center (corresponding to $2.8$ kpc after
correcting for inclination) along the major and minor axis,
respectively. The velocity dispersion declines from a central value of
about 190 to 150 \kms . Along the major axis we observe a
maximum $\Delta V \simeq 260$ \kms\ at $|r|\approx12''$.  Along the
minor axis there is no significant stellar velocity gradient and the
velocity dispersion ranges from a maximum of about 190 to 130 \kms\ at
the farthest measured radius.

The major-axis kinematics of the ionized gas is measured out to $57''$
($6.2$ kpc) from the nucleus. Gas rotates slower than stars in the
inner $|r|\la4''$ (0.4 kpc). The contrary is true further out. Ionized-gas
reaches its maximum $\Delta V \simeq 420$ \kms\ at $|r|\approx20''$.
At larger radii the gas rotation velocity decreases to about $100$
\kms\ on both sides of the nucleus. We interpret this velocity trend as
the result of a projection effect rather an intrinsic phenomenon and
therefore we conclude that also the gaseous disk tends to be more face
on at larger radii as well as the stellar disk. The gas velocity
dispersion peaks to a central value of about $150$ \kms, and outwards
it decreases remaining lower than $60$ \kms\ for $|r|\ga10''$.
We derived the minor-axis kinematics of the gaseous component out to
$11''$ ($1.5$ kpc after deprojecting) and to $41''$ ($5.7$ kpc after
deprojecting) from the NTT and ESO 1.52-m spectrum, respectively.
In the NTT spectrum the velocity curve shows a steep gradient in the
nucleus rising to a maximum observed rotation of about 75 \kms\ at
$|r|\approx2''$. At larger radii gas velocity drops to zero and no
rotation is detected for $|r|\ga6''$. The presence of a sharp rotation
of the gas component along the galaxy minor axis is confirmed by the
kinematics obtained from the ESO 1.52-m spectrum, although these data
are characterized by a lower spectral and spatial resolution with
respect to the NTT ones.  The gas velocity dispersion along the minor
axis shows the same behaviour observed along the major one.

%%%%%%%%%%%%%%%%%%%%%%%%%%%%% 
% DISCUSSION AND CONCLUSIONS 
%%%%%%%%%%%%%%%%%%%%%%%%%%%%% 
\section{Discussion and conclusions} 
\label{sec:n2855_conclusions} 

The analysis of the velocity curves of stars and ionized gas along
both the major and minor axis of NGC 2855 shows a kinematical
decoupling between the gas in the innermost regions of the galaxy with
respect to remaining gas.
Indeed the innermost gas shows a central steep velocity gradient along
the minor axis (for $|r|\la2''$) where stars do not have any
significant rotation. The gas rotates slightly slower than stars along
the galaxy major axis (for $|r|\la4''$). On the contrary, the outer
gas rotates faster than stars along the major axis and shows no
rotation along the galaxy minor axis.

We exclude that the gas velocity gradient we measure along the galaxy
minor axis is due to non-circular motions in a barred potential.
Indeed we do not observe any bar structure in low-inclined disk of the
galaxy neither on our $R-$band image nor in the near-infrared images
of Peletier et al. (1999) and M\"ollenhoff \& Heidt (2001). In
addition, the radial profiles of ellipticity and position angle do not
show the signatures of the presence of a nuclear bar, as discussed by
Wozniak et al. (1995).

Since the minor-axis gas rotation is measured inside the
bulge-dominated region and since the intrinsic shape of bulges is
generally triaxial (Bertola, Vietri \& Zeilinger 1991) we can consider
the kinematical decoupling observed in NGC 2855 as an effect due to
the bulge triaxiality.
If the gas component is in one the equilibrium planes of the bulge (as
in the case of the Sa NGC 4845 studied by Bertola, Rubin \& Zeilinger
1989; Gerhard, Vietri \& Kent 1989) the sharp rise and fall of the gas
rotation velocity observed for $|r|\la6''$ along the minor axis might
be caused by strong non-circular motions. However this is unlikely,
since we expect a mild triaxiality for the bulge of NGC 2855 because
of the lack of a significant isophotal twisting for $r\la10''$ (Fig.
\ref{fig:photometry}).
We therefore interpret the observed ionized-gas velocity field as due
to the presence of two kinematically-decoupled gaseous components,
which are rotating around two roughly orthogonal axes. They are the
shortest and the longest axes of the triaxial bulge. In this
picture the innermost gas is moving onto the equilibrium plane  
orthogonal to the equatorial one, where both the galaxy disk and 
outer gas component are settled. If this is the case, the kinematical 
decoupling is also consistent with an abrupt inner warp of the 
ionized-gas component.
 
Kinematical decoupling between two components of a galaxy suggests the
occurrence of a second event, so it is easy to explain the existence
of the innermost orthogonally-rotating gas as due to external material
acquired from a direction close to the bulge equilibrium plane
orthogonal to that of the galaxy disk. This scenario is supported by
the presence of the faint structure which surrounds the galaxy and is
elongated in a direction close to the galaxy minor axis. Part of the
gaseous material associated to this loop could be sinked towards the
galaxy nucleus sweeping away the gas of the disk and giving origin to
the decoupled component.
 
Although it involves a second event, the mechanism of formation of the
kinematically-decoupled component in NGC 2855 seems to be different
from that of NGC 4698 and NGC 4672, where the entire disk could be
formed in the acquisition process (Bertola \& Corsini 2000).  In fact
it has to be pointed out that in NGC 2855 the innermost gas has not
yet turned into stars, while it is partially and completely
transformed into stars in NGC 4698 and NGC 4672, respectively.
According to this result, although the low inclination of NGC 2855
does not preclude that its bulge could be an oblate spheroid
protruding out of the disk plane but we have to explain why star
formation did not occur in the inner orthogonally-rotating gaseous
component. Such a kinematically-decoupled gas may represent an early
stage toward the formation of an orthogonally-rotating stellar core.

The case of NGC 2855 gives further support to the mounting evidence
that in nearby galaxies even with an apparently undisturbed morphology
we find the signatures of on-going or recent acquisition phenomena if
we study them in detail. A better knowledge of second events in
nowadays galaxies is a key to understand such a kind of processes in
the early universe, when they were more frequent due to the higher
galaxy density.

\acknowledgements 
We acknowledge Ralf-J\"urgen Dettmar for the ESO 3.6-m narrow-band
imaging of NGC 2855. We are indebted to R. Bender and R. Saglia for
providing us with the FCQ package we used for measuring the stellar
kinematics. We are also grateful to D. Malin for the image of NGC
2855 we used in Fig. \ref{fig:malin}.

\end{document}